# PRESENT STATUS OF THE EXPERIMENT ON THE SEARCH FOR DARK PHOTONS BY MULTI-CATHODE COUNTER

A. Kopylov[1], I. Orekhov, V. Petukhov

Institute for Nuclear Research of RAS,

Moscow, Prospect of 60th Anniversary of October Revolution 7A, Russia

Abstract. In this paper we describe the apparatus used in this experiment and the procedure of data treatment. We give the latest experimental results and discuss the possibility to observe diurnal variations of the count rates due to rotation of the Earth. We outline our future plans.

## 1. Introduction

The present observation of the stellar dynamics shows that about 85% of matter is contained in a non luminous mass, so called "dark matter". The origin of this is still unknown. Another enigma is that dark matter is distributed by a spherical halo that is drastically different from distribution of baryonic matter. Because the latter is governed by pure gravitational interactions this difference is a hint that may be there is some collective interaction between the constituents of dark matter responsible for this difference. For these collective interactions to be possible one should assume that number density for the dark matter should be high and the interaction should be long range similar to electromagnetic interaction. As an ideal candidate to adopt these properties would be the waves of dark photons with a rest mass less than a few tens of eV. This constitutes the motivation for the search of dark photons as a cold dark matter. Dark photons have been suggested as a modification of electrodynamics by L. Okun [1], the physics of dark photons is outlined in a number of reviews, for example in [2]. In this paper we describe the experiment PHELEX (PHoton-ELectron EXperiment) with the aim to detect dark photons using a special device – multi-cathode counter.

This paper is structured as follows. In section 2 we describe the apparatus used in this experiment. In section 3 we give description of the data treatment developed for this experiment. Here we also describe the uncertainties of measurement. In section 4 we present the results obtained. In section 5 we discuss the possibility to observe diurnal variations of the count rates due to rotation of the Earth. In section 6 we outline our future plans and in section 7 we present the implications of the results obtained in this work.

---

[1] kopylov@inr.ru

## 2. Description of the apparatus

We have developed a new method as a further extrapolation of a method of dish-antenna [3] where photons are detected as a result of a conversion of dark photons on the surface of dish-antenna. At higher energies the coefficient of reflection of photons from the surface of dish-antenna is low and the photons are not getting reflected but getting absorbed by the surface. If the material of the surface is metal then electrons are emitted by certain quantum efficiency. In our experiment we are looking for single-electrons emitted from a surface of a metal cathode of a proportional counter with high (about $10^5$) gas amplification. We have described this method in our previous publications [4-6] here we give only some general information needed for understanding the experiment. In Fig.1 the scheme of a multi-cathode counter is presented.

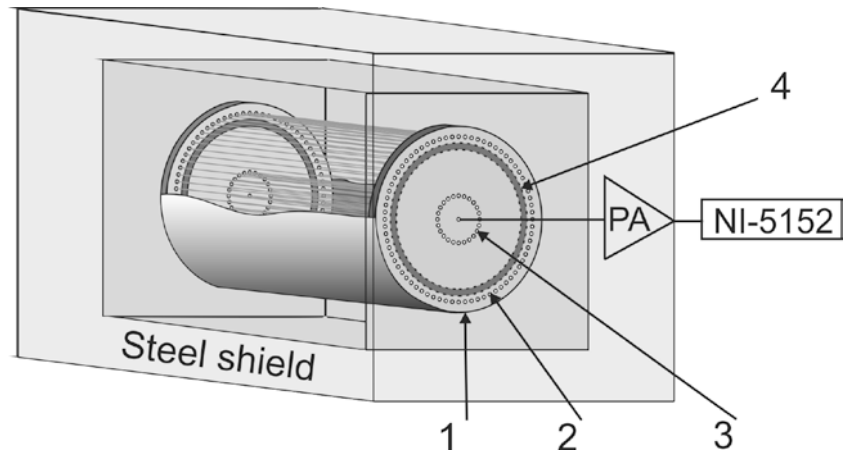

Fig.1. Scheme of a multi-cathode counter with electronic modules. 1 – cathode #1, 2 – cathode #2, 3 – cathode #3, 4 – focusing ring

The original feature of this detector is a special construction with multiple cathodes. First cathode is a cylinder made of metal 160 mm in diameter. The electrons are emitted from the surface of this cathode by the conversion of dark photons. These electrons freely drift to the anode in the center of the counter made of gold plated tungsten wire of 20 μm in diameter. Near this wire third cathode is arranged made of nichrome wires 50 μm in diameter. This cathode enables to get high ($> 10^5$) gas amplification needed for detection of the signal from single electron. In the vicinity of the first cathode at the distance of about 5 mm another (second) cathode is arranged also made from 50 μm nichrome wires. This cathode is used to control the drift of electrons from first cathode to the anode of the counter. When the potential applied to this cathode is higher than the potential applied to first cathode, the electrons emitted from first cathode freely drift towards an anode of the counter. This is first configuration when we count the events from the electrons emitted from first cathode plus background from other sources like traces of particles cut at the ends of counter. When the potential applied to this cathode is lower than the potential applied to first cathode, the electrons emitted from first cathode are rejected back and do not reach an anode of the counter. This is second configuration when only background events are counted. The real effect can be found by subtracting the count rate in the second configuration from the count rate obtained in first configuration. To reduce the background from tracks cut at the ends of the counter the focusing rings have been placed at both ends of the counter. They prevent the collection of the charge from ionizing particles by the isolating discs at both ends of the counter.

The counter was filled initially by Ar + $CH_4$(10%) and later by Ne + $CH_4$(10%) gas mixture. Using Ne turned out to be more advantageous because of higher gas amplification by lower fields. The counter was calibrated by UV lamp through a special quartz window in the wall.

The detector is placed in a cabinet with a steel shield to reduce the background from the external gamma-radiation. This resulted in the attenuation of the count rate in both configurations by a factor of two. The whole assembly is situated at sea level in Troitsk, Moscow. The signal from the anode of the counter was fed to the input of charge-sensitive preamplifier and then to the input of ADC NI-5152. The counting has been performed by 12 hours series: day- and night-counting. As a result more than 1 TB of data has been collected each day. The data treatment has been conducted off-line.

### 3. The procedure of data treatment

The detector has been placed in a shielding cabinet at sea level. The single electron pulses with an average amplitude of about 9 mV have a Polya distribution [5]. From this distribution it was taken that "useful" pulses belong to the region from 3 to 30 mV with the efficiency 57%. Figure 2 shows one of the captured frames where one can see big pulse from muon crossing a counter and a small pulse from the single electron. As one can see from this figure, the main problem in the analyses of the raw data was to select small pulses from the big ones from muons with the distorted baseline. The problem was also to evaluate correctly the real live time of measurements. That was achieved using the digitizing board. The important factor was also the noise. In our measurements it was about 2 mV. In future we are planning to reduce the noise by a factor of 2 using a Faraday cage.

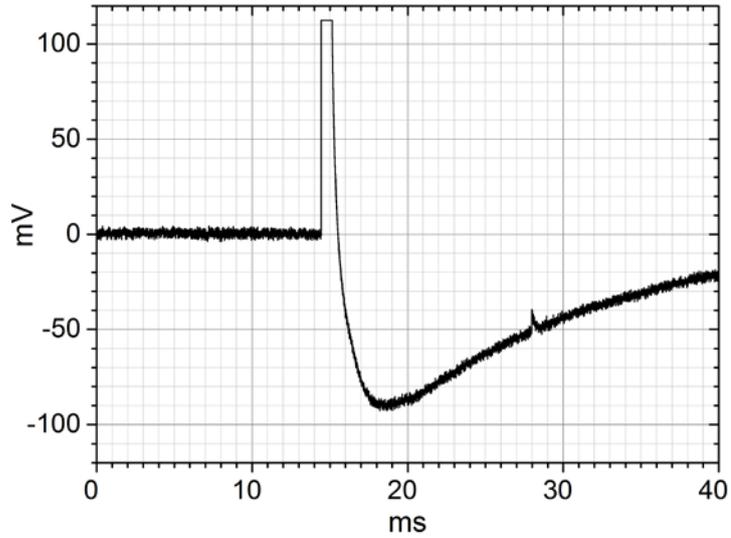

Fig.2. One of the captured frames

The interval 10 ms from the front edge of the big pulse has been discarded with a proper estimate of live-time. The amplitude of the small pulse has been restored from the real baseline observed. The final result was obtained by averaging the difference between count rates during counting in configuration 1 and the one counted in configuration 2. Figure 3 shows the results of measurement obtained with a counter filled by Ne + $CH_4$(10%) mixture.

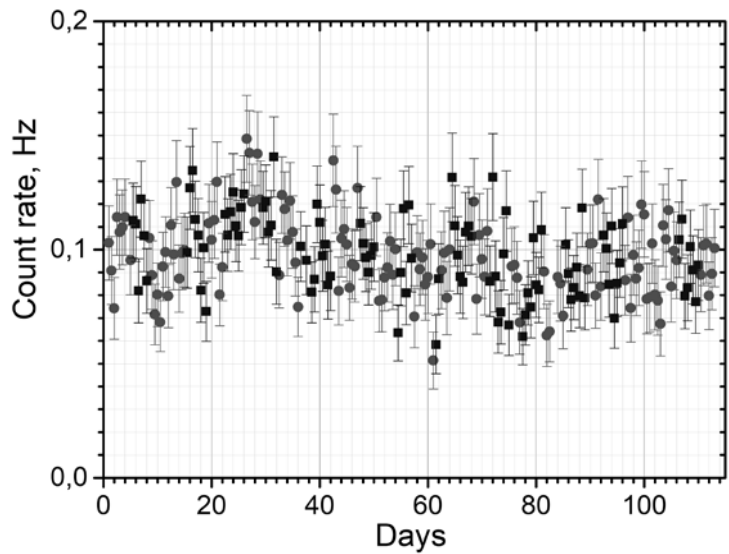

Fig.3. Count rates in configuration 1 (circles) and configuration 2 (squares).

## 4. The results obtained

The final results obtained in our measurements show that we have obtained the lowest count rate of single electrons emitted from metallic cathode (duralumin in this case) $r_{MCC} = (-0.33 \pm 0.7)\cdot 10^{-6}$ Hz/cm$^2$. From here we obtained an upper limit for a mixing parameter $\chi$ [7]. Figure 4 shows this result together with the ones obtained in other experiments.

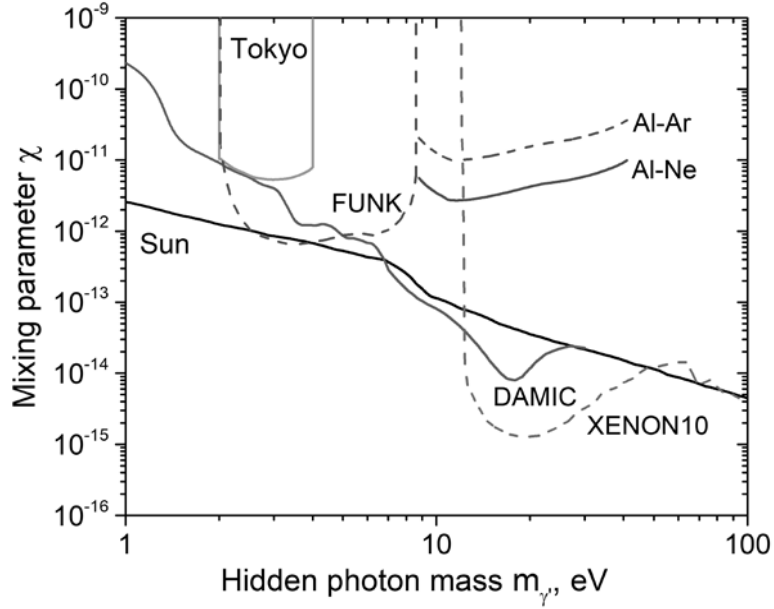

Fig4. The upper limits obtained. The results of other experiments: Tokyo [10], DAMIC [11], FUNK [12], XENON10 [13].

One should note that in our experiment the target is free electrons of a degenerate electron gas of a metal while in other experiment the target is valence electrons. The physics of these two cases is different in details and this may turn out to be very important. Hence, these experiments are complimentary to each other.

## 5. Diurnal variations of the count rates

The remarkable feature of our detector is that here the signal depends on the orientation of the counter if dark photons are polarized. The power absorbed by the cathode of the counter

$$P = 2\alpha^2 \chi^2 \rho_{CDM} A_{cath} \qquad (1)$$

here the average value $\alpha^2 = \langle \cos^2\theta \rangle$ and $\theta$ is the angle between the direction of the vector of the electric field and the surface of the cathode, $\rho_{CDM} = 0.46$ GeV/cm$^3$ – is the energy density of cold dark matter, $A_{cath}$ – the surface of the cathode. If the direction of the field is isotropic then $\alpha^2$

= 2/3. If electric field is parallel to the axis of the counter, the effect is zero. The maximum effect has achieved when the vector of the electric field is perpendicular to the axis. Then $\alpha^2 = \frac{1}{2}$. The rate of single-electrons is connected with the power collected by a cathode of the counter by

$$P = m_{\gamma'} R_{MCC}/\eta \qquad (2)$$

here $m_{\gamma'}$ is the mass (energy) of a dark photon, $R_{MCC}$ is the rate of emission of single-electrons and $\eta$ is the quantum efficiency for a photon with energy $m_{\gamma'}$ to yield a single-electron. If the field of dark photons is polarized, i.e. it has a certain orientation relative galactic or solar system then we should observe diurnal variations of the count rate due to rotation of the Earth. For the latter case the period of variations should be equal to 24 hours. For the former case the period should be equal to 23 hours 56 minutes. Principally it would be possible to find out which case is realized by collecting very high statistics. Remarkable thing is that the shape of the curve of diurnal variations depends also on the orientation of the counter and on the geographical latitude of the place where the detector is positioned [8]. Figures 5 shows the calculated diurnal variations for three orientations of the counter (Vertical, North-South and East-West) placed at three different sites with different geographical latitudes: Baksan Neutrino Observatory, Russia at 43ºN, INO in India at 10ºN and Pyhäsalmi mine in Finland at 64ºN. One can see that the curves are very characteristic and combination of all data obtained will enable to restore the vector of electric field in galactic or solar frame. But for this one should make the measurements with an array of counters during long time to obtain high statistics. One should also take into account that the polarization can be only partial so to extract the effect will be needed very high statistics.

BNO

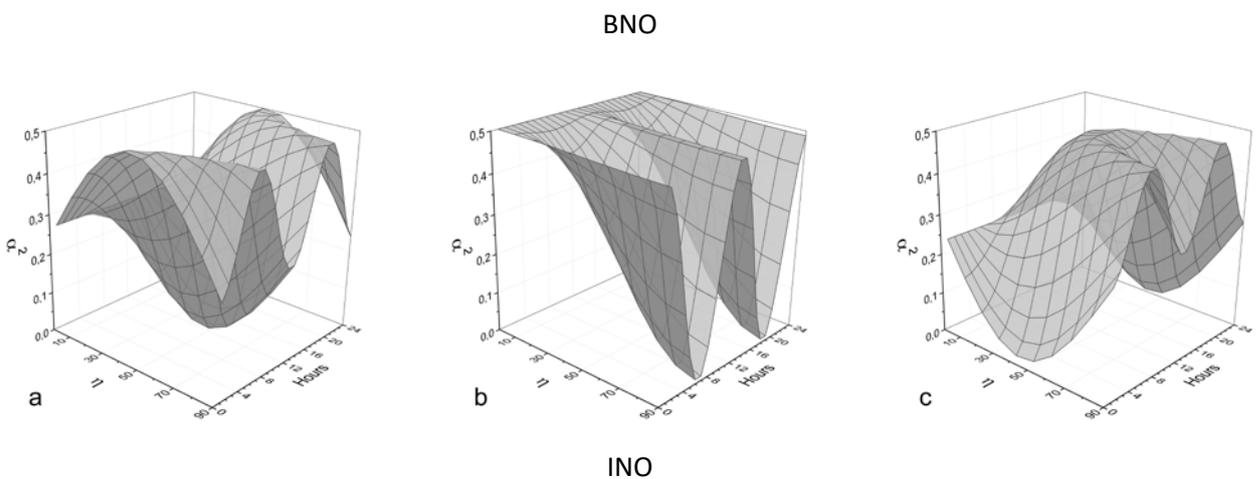

a  b  c

INO

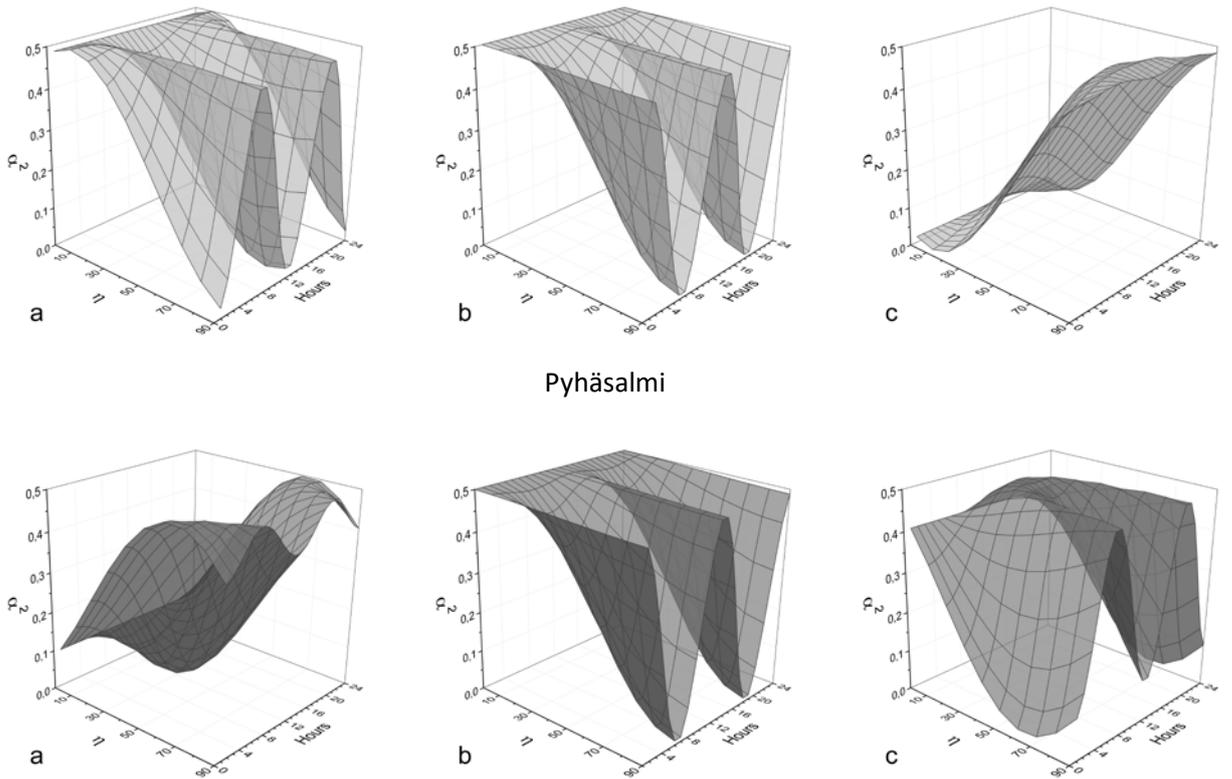

Fig.5. Averaged (for 1 hour) value $\alpha^2 = \langle \cos^2\theta \rangle$ as a function of time and angle $\eta$ (degrees) for three orientations of the counter: (a) – vertical one, (b) – along the parallel, (c) – along the meridian, each one - for three mines with different geographical latitudes.

### 6. Future plans

We are planning to improve our limit by further development of the counting system and by increasing the time of measurements to get higher statistics. To search for possible diurnal variations we are planning to conduct the measurements with an array of four counters: 3 of them with a mirror surface of a cathode at different orientation and one counter with a matt surface, when we should not observe diurnal variations because of averaging of the effect due to roughness of the cathode surface [9]. We want to place this array at Baksan Neutrino Observatory to decrease the effect from muons. Figure 6 shows the layout of this array of counters.

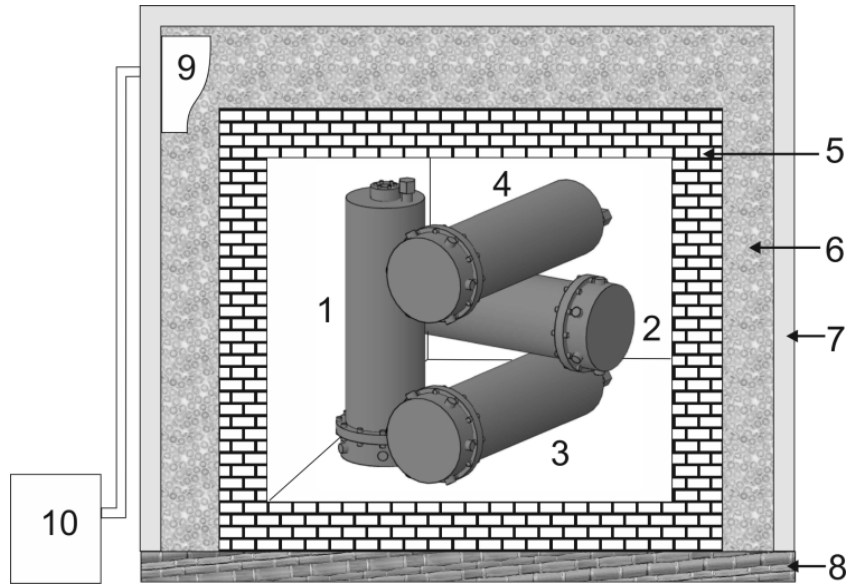

Fig.6. Array of counters placed in a shielded cabinet with a thermal insulation. 1 – 3 – counters in different orientation with a mirror surface, 4 – counter with a matt surface, 5 – lead bricks, 6 – thermal insulation, 7 – steel cabinet, 8 – wooden supporting plate, 9 – 10 – cooling system

They will be placed in a shielded cabinet with a thermal insulation and cooled by a special cooling system.

### 7. Conclusions

We have developed a novel technique to search for dark photons using a special detector: a multi-cathode counters. The method turned out to be very efficient for masses of dark photons from 9 to 40 eV, and first results are encouraging. We are planning to improve the results obtained and hope it will be possible to search for diurnal variations of the count rate by an array of multi-cathode counters. The great advantage of this method is that here the characteristic time variations of count rates obtained by an array of counters would be evidence that we really observe dark photons. If positive this study will enable to restore the vector of the electric field in solar or galactic frame.

We appreciate very much the substantial support from the Ministry of Science and Higher Education of the Russian Federation within the "Instrument base renewal program" in the framework of the State project "Science".